\documentclass[aps,showpacs, preprint, 12pt]{revtex4}
\usepackage{graphicx}
\usepackage{amsmath}
\usepackage{amsfonts}
\usepackage{amssymb} 
\usepackage[caption=false]{subfig}
\newcommand{\be}{\begin{equation}}
\newcommand{\ee}{\end{equation}}
\newcommand{\bel}[1]{\begin{equation}\label{#1}}
\newcommand{\bea}{\begin{eqnarray}}
\newcommand{\eea}{\end{eqnarray}}
\newcommand{\ba}{\begin{array}}
\newcommand{\ea}{\end{array}}

\begin{document}

\setlength{\unitlength}{1mm}

\title{ The totally asymmetric exclusion process with generalized update}
\author{A.E. Derbyshev}
\email{AndreyDerbishev@yandex.ru}
\author{S.S. Poghosyan}
\email{spoghos@theor.jinr.ru}
\author{A.M. Povolotsky}
\email{alexander.povolotsky@gmail.com}
\author{V.B. Priezzhev}
\email{priezzvb@theor.jinr.ru}

\affiliation{~\newline Bogolubov Laboratory of Theoretical Physics,
\\Joint Institute for Nuclear Research,
\\141980 Dubna, Russia}

\bigskip
\begin{abstract}
We consider the totally asymmetric exclusion process in discrete time with generalized updating rules.
We introduce a control parameter into the interaction between particles. Two particular values of the parameter
correspond to known parallel and sequential updates. In the whole range of its values the interaction varies from repulsive to attractive. In the latter case the particle flow demonstrates an apparent jamming tendency not typical for the known updates. We solve the master equation for $N$ particles on the infinite lattice by the Bethe ansatz.
The non-stationary solution for arbitrary initial conditions is obtained in a closed determinant form.
\end{abstract}
\pacs{05.40.+j, 02.50.-r, 82.20.-w}
\maketitle

\section{Introduction}

The totally asymmetric simple exclusion process (TASEP) is a paradigmatic model of stochastic systems of interacting particles
demonstrating  non-equilibrium behavior \cite{Spoh91,Ligg99,Gunter}.
The discrete-time dynamics of the model on the integer one-dimensional lattice is characterized by one of possible updating rules.
The most important cases are the backward-sequential, parallel, sublattice-parallel and random sequential updates \cite{Raje98}.
The first two of them are investigated in more detail than others. For a finite number of particles,
the dynamics can be defined through a master equation
\be
P(X,t+1) = \sum_{X^{'}} T(X,X^{'})P(X^{'},t)
\label{1}
\ee
where components of $N$-dimensional vector $X = \{x_i\},(x_1<x_2<\dots<x_N)$  are positions of $N$ particles and $T(X,X^{'})$
is the transition probability to go in one time step from configuration $X^{'}$ to configuration $X$.  For the backward-sequential
update, each particle may take one step to the right with probability $p$
if the target site is vacant at the beginning of the time step or becomes vacant at the end of the time step due to motion of the particle
in front. For the parallel update, the motion to the right is allowed only if the target site is vacant at the beginning of the time step.
By iterating (\ref{1}) one obtains the solution of the master equation for any given
initial configuration $X^0$, that is the conditional probability to find a particle configuration $X$ at time step $t$,
given that the process started from configuration $X^0$.

A common property of the above mentioned updating rules is the short range repulsion between particles due to the exclusion condition.
On large scales, the particle density evolves according to the Burgers equation \cite{Rost}. Fluctuations of the density and current on an appropriate scale are the subject of intensive studies \cite{Derrida_rev,PS1,BC} over the past two decades.
The second important property of the mentioned updates is their "solvability" which means that the master equation (\ref{1}) can be solved
by the Bethe ansatz method and the Green functions can be calculated explicitly \cite{Gwa,Schutz}.
In this paper, we propose a generalization of updating rules in order to modify the first property retaining the second one. Namely,
besides the exclusion condition, we allow an attractive interaction between neighboring particles which changes evolution of the system drastically.

Specifically, we introduce a control parameter into the interaction between particles. Depending on its value, we obtain an attraction or an additional repulsion between particles. We consider the TASEP on the infinite one-dimensional lattice and solve the master equation for arbitrary initial conditions to find the Green function in the determinant form as it was done in \cite{PriezBrank,pov pri,PogPriS}.
The solution generalizes the known results for the backward-sequential and parallel updates and coincides
with them for particular values of the parameter of interaction.

The article is organized as follows. In the Section II, we formulate the model and define basic notions needed for the solution.
In the Section III, we explain a peculiarity of the master equation in our case and transform the evolution operator to a form
appropriate for the Bethe ansatz. Solving the master equation for arbitrary initial conditions, we obtain the closed expression
for the Green function. The Appendix contains the proof of the Bethe ansatz in the general $N$-particle case.

\section{Definitions}

\begin{figure}[tbp]
\includegraphics[width=0.6\textwidth]{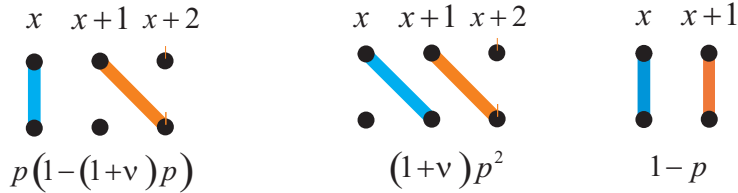}
\caption{The probabilities of different jumps of two neighboring particles. }
\label{Fig1}
\end{figure}

Consider $N$ particles moving to the right on the one-dimensional lattice.
At each discrete moment of time, each lattice site can be occupied by at most one particle.
To describe the dynamics of the model, we follow the backward-sequential rule: at given time step, we scan the particle configuration
from the right to the left and consider situations when two particles meet in neighboring sites just before the time step (see Fig.\ref{Fig1}).
If the right particle of the pair does not hop at given step,
the left particle stays at its position with probability $1$. If the right particle hops, the next particle
jumps to the right with probability $(1+\nu)p$,
or stands with probability $1-(1+\nu)p$. A natural restriction on the parameter $\nu$ is $0 \leq (1+\nu) \leq 1/p$.
For $\nu=0$, we have the  usual TASEP with backward-sequential update. For $\nu=-1$, we obtain the TASEP with parallel update.

\begin{figure}[tbp]
 \subfloat{\includegraphics[width=0.45\textwidth]{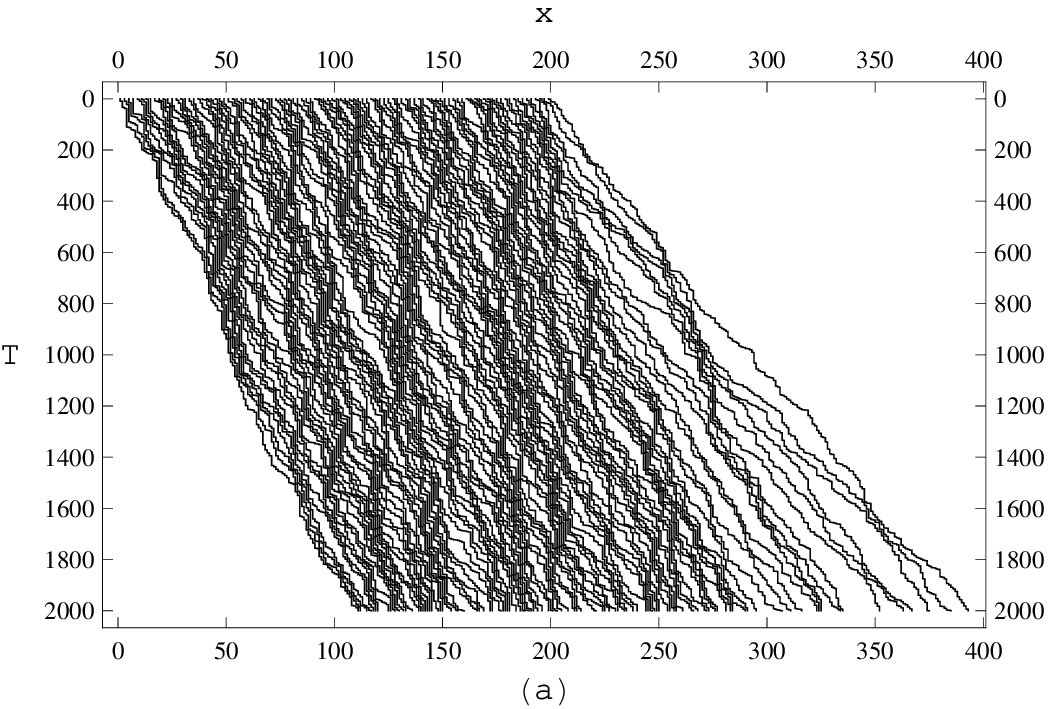}}~~~
\subfloat{\includegraphics[width=0.45\textwidth]{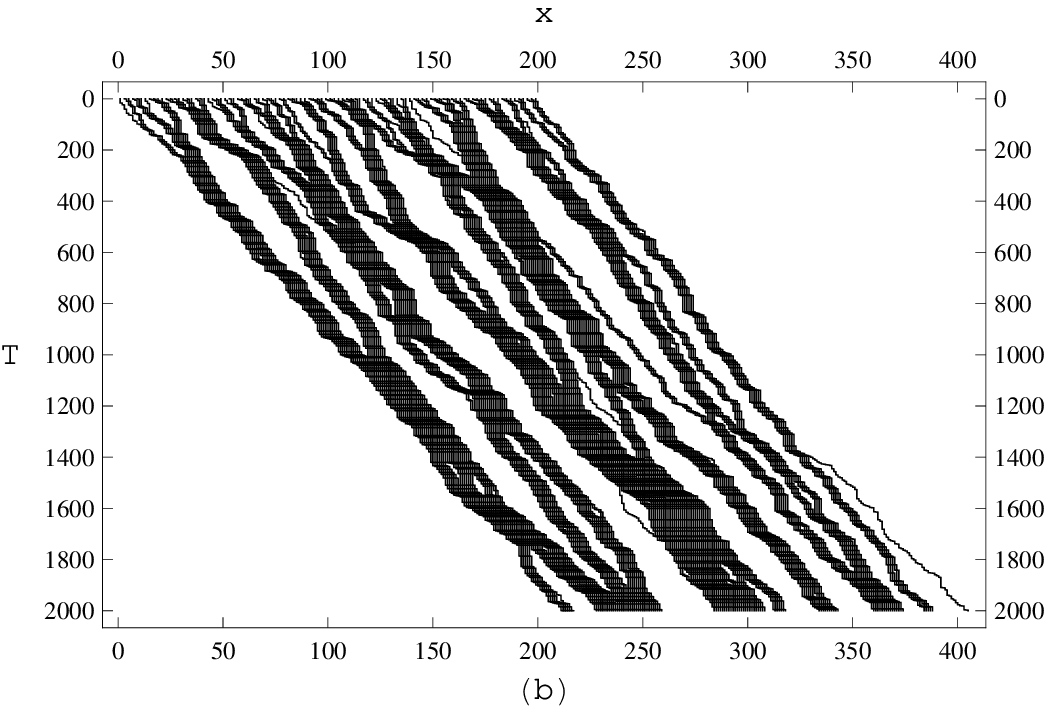}}
\caption{Typical behaviour of  particle space-time trajectories at (a) rarefaction regime, with $p=0.1$, $\nu=-1$, and (b) jamming regime, $p=0.1$ and $\nu=8.95$. The dynamics of $N=100$ particles is simulated for $T=2000$ time steps starting with the initial configuration where particles are alternating with empty sites. }
\label{Fig2}
\end{figure}

If $\nu > 0$, the system demonstrates a collective behavior which is quite typical for the car traffic. To see this, consider
a limiting illustrative example $\nu=1, p=1/2$. Isolated particles move forward as usual with the probability $p=1/2$. If two particles
turn out nearest neighbors, they become "stuck together" and move synchronously because the situation when the right particle moves and the
left one stands is forbidden. This micro-jam grows involving into oneself new and new particles until the moment when all particles move synchronously in a huge $N$-particle jam.
Thus, when varying the parameter $\nu$   from $-1$ to $(1/p-1)$ the system passes through a qualitative change of particle density behaviour: from rarefaction to clumping. Two typical examples of such behaviour are shown in  Fig.(\ref{Fig2})

We start solving the problem with several definitions.
We introduce the Hilbert space supplied with the complete left and right
bases consisting of vectors $\left\langle X\right\vert $ and $\left\vert
X\right\rangle $ respectively, where $X$ runs over all particle
configurations, with inner product
\begin{equation}
\left\langle X|X^{\prime }\right\rangle =\delta (X,X^{\prime }).
\end{equation}%
The state of the system at any time step can be associated with the state
vector%
\begin{equation}
\left\vert P_{t}\right\rangle =\sum_{\left\{ X\right\}
}P_{t}(X)\left\vert X\right\rangle .
\end{equation}%
In terms of the state vectors, the master equation (\ref{1}) takes a simple operator
form
\begin{equation}
\left\vert P_{t+1}\right\rangle =\mathbf{T}\left\vert P_{t}\right\rangle ,
\end{equation}%
where the evolution operator $\mathbf{T}$ is defined by the transition probabilities%
\begin{equation}
\mathbf{T=}\sum_{\{X\},\{X^{\prime }\}}\left\vert X\right\rangle
T(X,X^{\prime })\left\langle X^{\prime }\right\vert .
\end{equation}%
The conditional probability $P(X,t|X^{0},0)$ can be represented as the
matrix element
\begin{equation}
P(X,t|X^{0},0)=\left\langle X\right\vert \mathbf{T}^{t}\left\vert
X^{0}\right\rangle .  \label{T^t}
\end{equation}%
To evaluate matrix elements, we construct the set of left eigenvectors $%
\left\vert B_{Z}\right\rangle $ of the operator $\mathbf{T}$%
\begin{equation}
\mathbf{T}\left\vert B_{Z}\right\rangle =\Lambda (Z)\left\vert
B_{Z}\right\rangle  \label{T|Z>=Lambda(Z)|Z>}
\end{equation}%
and the adjoint set of right eigenvectors $\left\langle \overline{B}_{Z}\right\vert $
\begin{equation}
\left\langle \overline{B}_{Z}\right\vert \mathbf{T}=\Lambda (Z)\left\langle
\overline{B}_{Z}\right\vert ,  \label{<z|T=lambda<z|}
\end{equation}%
where $Z$ is a $N$-dimensional parameter.
Now, the transition probability (\ref{T^t}) can be reduced to the evaluation of the sum
\begin{equation}
\left\langle X\right\vert \mathbf{T}^{t}\left\vert X^{0}\right\rangle
=\sum\limits_{Z}\left\langle X|\mathbf{T}^{t}|B_{Z}\right\rangle
\left\langle \overline{B}_{Z}|X^{0}\right\rangle =\sum\limits_{Z}\Lambda
^{t}\left( Z\right) \left\langle X|B_{Z}\right\rangle \left\langle \overline{%
B}_{Z}|X^{0}\right\rangle   \label{<x|T^t|x_0>},
\end{equation}
where the range of summation is to be defined from the boundary conditions.
\section{Solution}

To understand a peculiarity of the eigenproblem (\ref{T|Z>=Lambda(Z)|Z>}) in our case,
we consider first the two-particle problem. If two particles are at least two empty sites apart from each
other, they move independently and Eq.(\ref{T|Z>=Lambda(Z)|Z>}) is equivalent to the equation
\begin{eqnarray}
\Lambda \left( Z\right) \left\langle x_{1}
,x_{2}|B_{Z}\right\rangle &=&\sum_{\left\{k_1,k_2 \right\}
}\prod_{i=1}^{2}p^{k_{i}}\left( 1-p\right) ^{1-k_{i}}  \notag \\
&&\times \left\langle x_{1}-k_{1},x_{2}-k_{2}|B_{Z}\right\rangle
.  \label{masterfree}
\end{eqnarray}%
where the summation is over $k_1,k_2$ which take values 1 or 0 depending on whether the particle decided to
jump or not. If particles are neighbors, $x_1+1=x_2$, the terms in the RHS of the actual (interacting) equation \eqref{T|Z>=Lambda(Z)|Z>} are
$ p^2(1+\nu)\left\langle x_{1}-1,x_{2}-1|B_{Z}\right\rangle$, $p(1-p)\left\langle x_{1}-1,x_{2}|B_{Z}\right\rangle$,
$(1-p)\left\langle x_{1},x_{2}|B_{Z}\right\rangle$,  whereas the term $p(1-p)\left\langle x_{1},x_{2}-1|B_{Z}\right\rangle=
p(1-p)\left\langle x_{1},x_{1}|B_{Z}\right\rangle$ in the free equation (\ref{masterfree}) is beyond the set of allowed configurations. The strategy of the Bethe ansatz is to define formally the forbidden term via allowed terms in such a way that Eq.(\ref{T|Z>=Lambda(Z)|Z>})
retains its "free" form (\ref{masterfree}) in the case $x_1+1=x_2$. Then the formal relation between forbidden and allowed terms can be used for determination of coefficients of the Bethe substitution
\begin{equation}
 \left\langle x_{1},x_{2}|B_{Z}\right\rangle = A_{12}z^{-x_1}_1 z^{-x_2}_2 + A_{21} z^{-x_1}_2 z^{-x_2}_1.
\label{Bethe}
\end{equation}

This strategy is effective for the TASEP with the backward-sequential update. In our case (as well as in the case of
parallel update \cite{pov pri}) an additional problem arises. To see it, consider configuration $\{x_1,x_2\}$ where
$x_1+2=x_2$. The terms in the RHS of  Eq.(\ref{T|Z>=Lambda(Z)|Z>}) are $ p^2\left\langle x_{1}-1,x_{2}-1|B_{Z}\right\rangle$,
$p(1-p)\left\langle x_{1}-1,x_{2}|B_{Z}\right\rangle$,
$(1-p)^2\left\langle x_{1},x_{2}|B_{Z}\right\rangle$ and $p(1-(1+\nu)p)\left\langle x_{1},x_{2}-1|B_{Z}\right\rangle$.
The last term differs from its free version $p(1-p)\left\langle x_{1},x_{2}-1|B_{Z}\right\rangle$ and we have no
forbidden terms like $p(1-p)\left\langle x_{1},x_{1}|B_{Z}\right\rangle$ in the previous paragraph, to compensate this
difference. Instead, we can notice that the configuration $\{x_1,x_2-1\}$ contains the pair of neighboring particles,
which disappears in the configuration $\{x_1,x_2\}$. Thus, if we multiply the transition probability $T(X,X^{'})$
by the factor $\lambda = (1-p)/(1-(1+\nu)p)$ corresponding to dissociation of the pair, we restore the free form
of the last term in the RHS of (\ref{masterfree}). In the $N$-particle case, the factor $\lambda$ should be ascribed
to each dissociated pair.

More formally, we define an auxiliary matrix
\begin{equation*}
\mathbf{T}_{0}=\mathcal{D}\mathbf{T}\mathcal{D}^{-1}
\end{equation*}%
where $\mathcal{D}$ is a diagonal operator
\begin{equation*}
\mathcal{D}=\sum\limits_{\{X\}}\frac{\left\vert X\right\rangle \left\langle
X\right\vert }{W(X)}.
\end{equation*}%
and $W(X)$ depends on the number of pairs $N_p(X)$ in configuration $X$: $W(X)=\lambda^{N_p(X)}$.

Operators $\mathbf{T}$ and $\mathbf{T}_{0}$ have the same eigenvalues $\Lambda_Z$ and the right eigenvectors are related
by
\begin{equation}
\left\vert B_{Z}\right\rangle =\mathcal{D}^{-1}\left\vert
B_{Z}^{0}\right\rangle .  \label{<x|B_z>}
\end{equation}%
The matrix elements of $\mathbf{T}_{0}$ are
\begin{equation}
T_{0}(X,X^{\prime })  =\frac{W(X^{\prime })}{W(X)}
 T(X,X^{\prime }).
\label{T=T W(x')/W(x)}
\end{equation}%
The two-particle eigenproblem for operator $\mathbf{T}_{0}$ has the free form
\begin{eqnarray}
\Lambda \left( Z\right) \left\langle x_{1}
,x_{2}|B_{Z}^0\right\rangle &=&\sum_{\left\{k_1,k_2 \right\}
}\prod_{i=1}^{2}p^{k_{i}}\left( 1-p\right) ^{1-k_{i}}  \notag \\
&&\times \left\langle x_{1}-k_{1},x_{2}-k_{2}|B_{Z}^0\right\rangle.  \label{modmasterfree}
\end{eqnarray}%
if the distance between particles exceeds 1. If particles are neighbors, $x_1+1=x_2$, the terms in the RHS of (\ref{modmasterfree}) are
$ p^2(1+\nu)\left\langle x_{1}-1,x_{2}-1|B_{Z}^0\right\rangle$, $p(1-p)/\lambda\left\langle x_{1}-1,x_{2}|B_{Z}^0\right\rangle$,
$(1-p)\left\langle x_{1},x_{2}|B_{Z}^0\right\rangle$ and the term $p(1-p)\left\langle x_{1},x_{2}-1|B_{Z}^0\right\rangle$ in the
free equation is forbidden. Imposing the condition
\begin{eqnarray}
\langle x,x|B^0_Z\rangle=\frac{{\nu}p}{1-p}(\langle x-1,x|B^0_Z\rangle-\langle x-1,x+1|B^0_Z\rangle)+\langle x,x+1|B^0_Z\rangle.
\label{condition}
\end{eqnarray}
on the forbidden term, we convert the remaining three terms into the free form, ensuring correctness of (\ref{modmasterfree})
for arbitrary allowed configurations $\{x_1,x_2\}$.

In Appendix, we prove that (\ref{condition}) written for the general case as
\begin{equation}
\begin{array}{l}
\langle\dots,x,x,\dots|B^0_Z\rangle=\frac{{\nu}p}{1-p}(\langle\dots,x-1,x,\dots|B^0_Z\rangle-\langle\dots,x-1,x+1,\dots|B^0_Z\rangle)\\
\;\;\;\;\;\;\;\;\;\;\;\;\;\;\;\;\;\;\;\;\;\;\;\;\;\;\;\;\;\;\;\;\;\;\;\;\;\;\;\;
\;\;\;\;\;\;\;\;\;\;\;\;\;\;\;\;\;\;\;\;\;\;\;\;\;\;\;\;\;\;\;\;\;\;\;\;\;\;\;\;
+\langle\dots,x,x+1,\dots|B^0_Z\rangle
\end{array}
\label{gencondition}
\end{equation}
is sufficient to convert  the general $N$-particle problem into the free form
\begin{eqnarray}
\Lambda \left( Z\right) \left\langle x_{1},\ldots
,x_{N}|B_{Z}^{0}\right\rangle &=&\sum_{\left\{ k_{i}\right\}
}\prod_{i=1}^{N}p^{k_{i}}\left( 1-p\right) ^{1-k_{i}}  \notag \\
&&\times \left\langle x_{1}-k_{1},\ldots ,x_{N}-k_{N}|B_{Z}^{0}\right\rangle.  \label{Nmodmasterfree}
\end{eqnarray}%
After that we can look for $B_{Z}^{0}\left( x_{1},\ldots ,x_{N}\right) $ in
the form of the Bethe ansatz,%
\begin{equation}
\left\langle x_{1},\ldots ,x_{N}|B_{Z}^{0}\right\rangle =\sum_{\left\{
\sigma \right\} }A_{\sigma _{1}\ldots \sigma _{N}}z_{\sigma
_{1}}^{-x_{1}}\ldots z_{\sigma _{N}}^{-x_{N}},  \label{Bethe ansatz}
\end{equation}%
where parameter $Z$  is a set of $N$\ complex numbers $%
Z=\left\{ z_{1},\ldots ,z_{N}\right\} $. The summation is
over all permutations $\sigma $ of numbers $1,\ldots ,N$. \ The
substitution of (\ref{Bethe ansatz}) to the equation (\ref{Nmodmasterfree})
leads to the eigenvalues
\begin{equation}
\Lambda \left( Z\right) =\prod\limits_{i=1}^{N}\left(
1-p+ p z_{i}\right) ,  \label{Lambda(Z)}
\end{equation}%
and the substitution of (\ref{Bethe ansatz}) into the constraint (\ref{gencondition})
gives the relation between the amplitudes $A_{\sigma _{1}\ldots \sigma
_{N}}$, which differ from each other only by permuted indices %
\begin{equation}
\frac{A_{\ldots ij\ldots }}{A_{\ldots ji\ldots }}= - \frac{1-1/z_{i}}{1-1/z_{j}}\frac{1-\gamma z_{j}}{1-\gamma z_{i}}.
\label{A_ji=A_ij...}
\end{equation}%
where $\gamma=1-1/\lambda$. Separately, amplitudes  $A_{\sigma _{1}\ldots \sigma_{N}}$ are
\begin{equation}
A_{\sigma _{1}\ldots \sigma _{N}}=\left( -1\right) ^{\mathcal{P}\left(
\left\{ \sigma \right\} \right) }\prod\limits_{i=1}^{N}\left( -\frac{%
1-\gamma z_{\sigma _{i}}}{1-1/z_{\sigma _{i}}}\right) ^{i-\sigma _{i}},
\label{A_sigma}
\end{equation}%
where $\mathcal{P}\left( \left\{ \sigma \right\} \right) $ is the parity of
a permutation $\left\{ \sigma \right\} $.
Then, we are in position to write the element $\left\langle X|B_{Z}\right\rangle $ in (\ref{<x|T^t|x_0>}).

\begin{figure}[tbp]
\includegraphics[width=0.5\textwidth]{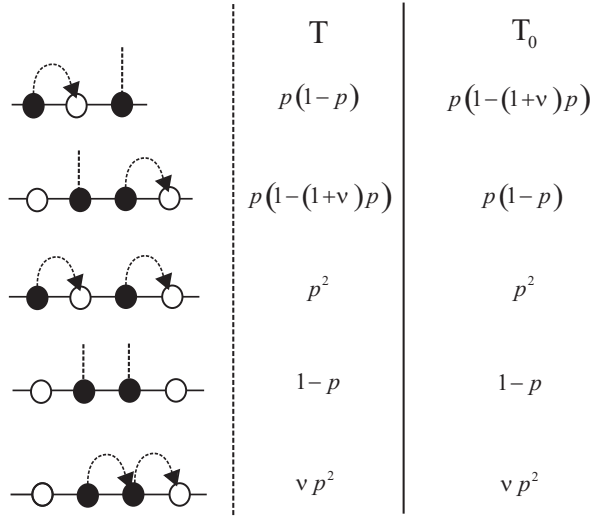}
\caption{The transition probability for the processes defined by  operators $\mathbf{T}$ and $\mathbf{T_0}$. }
\label{Fig3}
\end{figure}

To find $\left\langle \overline{B}_{Z}|X^{0}\right\rangle$, we examine the right eigenproblem (\ref{<z|T=lambda<z|}).
As above, we start with the two-particle case. If $x_2-x_1 \geq 2$, Eq.(\ref{<z|T=lambda<z|}) is equivalent to the equation
\begin{eqnarray}
\Lambda \left( Z\right) \left\langle \overline {B}_Z|x_{1},x_{2}\right\rangle &=&\sum_{\left\{k_1,k_2 \right\}
}\prod_{i=1}^{2}p^{k_{i}}\left( 1-p\right) ^{1-k_{i}}  \notag \\
&&\times \left\langle \overline{B}_Z|x_{1}+k_{1},x_{2}+k_{2}\right\rangle.  \label{rightmasterfree}
\end{eqnarray}
where the summation is over $k_1,k_2$ taking values 1 or 0. If $x_2-x_1 =1$, the RHS of (\ref{<z|T=lambda<z|})  contains three
terms: $p^2(1+\nu)\left\langle \overline{B}_Z|x_{1}+1,x_{2}+1\right\rangle$, $ p(1-(1+\nu)p)\left\langle \overline{B}_Z|x_{1},x_{2}+1\right\rangle$ and
 $ (1-p)\left\langle \overline{B}_Z|x_{1},x_{2}\right\rangle$. At the same time, the RHS of Eq.(\ref{rightmasterfree}) contains four terms  in this case, and one of them, $p(1-p)\left\langle \overline{B}_Z|x_{1}+1,x_{2}\right\rangle$,
 is forbidden. We can see from Fig.\ref{Fig3}

that  the right eigenproblem for the operator $\mathbf{T}$ in the two-particle case is similar, up to the inversion of the coordinates, to the left eigenproblem of the operator $\mathbf{T}_{0}$. This is true for the general $N$-particle
case and, therefore, we can use for $\left\langle \overline{B}_{Z}|X\right\rangle$ the similar Bethe ansatz:
\begin{equation}
\left\langle \overline{B}_{Z}|X\right\rangle =\frac{1}{N!}\sum_{\left\{
\sigma \right\} }\overline{A}_{\sigma _{1}\ldots \sigma _{N}}z_{\sigma
_{1}}^{x_{1}}\ldots z_{\sigma _{N}}^{x_{N}}  \label{<z|x>}
\end{equation}%
with the amplitudes $\overline{A}_{\sigma }$ given by
\begin{equation*}
A_{\sigma _{1}\ldots \sigma _{P}}=1/\overline{A}_{\sigma _{1}\ldots \sigma
_{N}}
\end{equation*}%
 Collecting obtained expressions, we have
\begin{eqnarray}
\left\langle X|B_{Z}\right\rangle  &=&W(X)\det \mathbf{B,} \\
\left\langle \overline{B}_{Z}|X\right\rangle  &=&\frac{1}{N!}\det \overline{%
\mathbf{B}}
\label{B&B}
\end{eqnarray}%
where the matrix elements $B_{ij}$ and $\overline{B}_{ij}$ of the matrices $%
\mathbf{B}$ and $\overline{\mathbf{B}}$ are%
\begin{equation}
B_{ij}=1/\overline{B}_{ij}=\left( -\frac{1-\gamma z_{j}}{1-1/z_{j}}\right)
^{i-j}z_{j}^{-x_{i}}.  \label{B_ij}
\end{equation}
The factor $1/N !$ is a normalization constant which follows from the condition
\begin{equation}
\sum_{Z}\left\langle X|B_{Z}\right\rangle \left\langle \overline{B}_{Z}|X\right\rangle = 1.
\label{<x|x>}
\end{equation}%
The range of summation over $Z$ is defined from the boundary conditions. For the case of
the infinite lattice treated here, the spectrum is continuous,
such that  the sum in
(\ref{<x|T^t|x_0>}) can be written as  integral
\begin{equation}
\sum\limits_{Z}\rightarrow \prod\limits_{i=1}^{N}\oint_{\Gamma_{0,1}} \frac{dz_{i}}{2\pi
iz_{i}},
\end{equation}%
where each $z_i$ runs along the closed contour ${\Gamma_{0,1}}$ in complex plane, which encircles the points $z=0,1$
 and leaves all  other singularities outside. The choice of the contour is justified by the completeness requirement
in a sense that the set of operators  $\{\left|B_{Z}\right\rangle \left\langle \overline{B}%
_{Z}\right|: Z\in \Gamma_{0,1}^{ N}\}$ provides the resolution of the identity operator
\begin{equation}
\sum\limits_{Z}\left\langle X|B_{Z}\right\rangle \left\langle \overline{B}%
_{Z}|X^{\prime }\right\rangle =\left\langle X|X^{\prime }\right\rangle
\label{<x|x'>},
\end{equation}%
which has been proved in \cite{pov pri} for the case of parallel update. The proof in
the general case differs by technical details only.

An explicit expression for the conditional probability $P(X,t|X^{0},0)$ follows from sequential substitutions
(\ref{B_ij}) into (\ref{B&B}), then (\ref{B&B}) and (\ref{Lambda(Z)}) into (\ref{<x|T^t|x_0>}) and (\ref{T^t}).
\begin{equation}
\left\langle X\right\vert \mathbf{T}^{t}\left\vert X^{0}\right\rangle
=W(X)\sum_{\left\{\sigma  \right\} }\left( -1\right) ^{\mathcal{P}\left( \left\{ \sigma \right\} \right)
}\prod\limits_{i=1}^{N}F_{i-\sigma_i}(x_i-x^0_{\sigma_i},t),
\label{<x|T^t|x_0>-explicit form}
\end{equation}%
where the function $f$ is
\begin{equation}
F_n(x,t)= \oint_{\Gamma_{0,1}}\frac{dz}{2\pi iz}\left(1-p+pz\right)^t\left(\frac{1-\gamma z}{1-1/z}\right)^n z^{-x} .
\end{equation}
Finally, we obtain the Green function as the determinant of matrix $\mathbf{M}$:
\begin{equation}
P(X,t|X^0,0)=\lambda^{N_p(X)}\det\mathbf{M},
\end{equation}
with matrix elements
\begin{equation}
M_{ij}=F_{i-j}(x_i-x^0_{j},t).
\end{equation}

In the following part of this work, we will use the obtained Green function for calculation of correlation functions
and probability distribution of the position of the last particle which is the Tracy-Widom distribution \cite{Johansson,rakos schutz}
in the case of standard updates.

\acknowledgements
This work was supported by the RFBR grant 06-01-00191a, the grant of the Heisenberg-Landau program and the DFG grant RI 317/16-1.

\appendix

\section{induction}

In this section,  using an induction we prove that using (\ref{gencondition}) one reduces the problem with interaction to the free equation  (\ref{Nmodmasterfree}). In fact below we give the proof only for the particular case of a single isolated cluster of particles. When there are several clusters which are more than one site apart from each other, they are independent, and this proof is applied to each clusters separately. In principle one has to consider also the case of clusters separated by single empty site. However technically this proof is similar to the one below and therefore we omit it.

We start with the  eigenproblem (\ref{Nmodmasterfree}) for three particles situated in the cluster $x_1=x, x_2=x+1, x_3=x+2$:
\begin{eqnarray}
\Lambda \left( Z\right) \left\langle x_{1},x_{2}
,x_{3}|B_{Z}^{0}\right\rangle &=&\sum_{\left\{ k_{i}\right\}
}\prod_{i=1}^{3}p^{k_{i}}\left( 1-p\right) ^{1-k_{i}}  \notag \\
&&\times \left\langle x_{1}-k_{1}, x_{2}-k_{2} ,x_{3}-k_{3}|B_{Z}^{0}\right\rangle.  \label{3modmasterfree}
\end{eqnarray}%
We have to prove that condition (\ref{gencondition}) converts (\ref{3modmasterfree}) into the eigenproblem of operator $\mathbf{T}_{0}$.
The RHS of (\ref{3modmasterfree}) contains four forbidden terms
$\langle x,x+1,x+1|B^0_Z\rangle$,  $\langle x-1,x+1,x+1|B^0_Z\rangle$, $\langle x,x,x+2|B^0_Z\rangle$ and $\langle x,x,x+1|B^0_Z\rangle$.
Using a straightforward extension of (\ref{condition}), we define two of them as
\begin{equation}
\langle x,x,x+2|B^0_Z\rangle=\frac{\nu p}{1-p}(\langle x-1,x,x+2|B^0_Z\rangle-\langle x-1,x+1,x+2|B^0_Z\rangle)+\langle x,x+1,x+2|B^0_Z\rangle
\label{forbidden_1}
\end{equation}
\begin{equation}
\langle x-1,x+1,x+1|B^0_Z\rangle=\frac{\nu p}{1-p}(\langle x-1,x,x+1|B^0_Z\rangle-\langle x-1,x,x+2|B^0_Z\rangle)+\langle x-1,x+1,x+2|B^0_Z\rangle
\label{forbidden_2}
\end{equation}
For  $\langle x,x+1,x+1|B^0_Z\rangle$ and $\langle x,x,x+1|B^0_Z\rangle$ we have the linear system
\begin{eqnarray*}
\begin{cases}
\langle x,x+1,x+1|B^0_Z\rangle=\frac{\nu p}{1-p}(\langle x,x,x+1|B^0_Z\rangle-\langle x,x,x+2|B^0_Z\rangle)+\langle x,x+1,x+2|B^0_Z\rangle\\
\langle x,x,x+1|B^0_Z\rangle=\frac{\nu p}{1-p}(\langle x-1,x,x+1|B^0_Z\rangle-\langle x-1,x+1,x+1|B^0_Z\rangle)+\langle x,x+1,x+1|B^0_Z\rangle
\end{cases}
\end{eqnarray*}
which gives
\begin{equation}
\langle x,x+1,x+1|B^0_Z\rangle=\frac{\nu^2p^2}{(1-p)^2}(\langle x-1,x,x+1|B^0_Z\rangle-\langle x-1,x,x+2|B^0_Z\rangle)
+\langle x,x+1,x+2|B^0_Z\rangle
\label{forbidden_3}
\end{equation}
and
\begin{equation}
\langle x,x,x+1|B^0_Z\rangle=\frac{{\nu}p}{1-p}(\langle x-1,x,x+1|B^0_Z\rangle-\langle x-1,x+1,x+2|B^0_Z\rangle)+\langle x,x+1,x+2|B^0_Z\rangle
\label{forbidden_4}
\end{equation}
Using definitions (\ref{forbidden_1}),(\ref{forbidden_2}), (\ref{forbidden_3}) and (\ref{forbidden_4}) we rewrite (\ref{3modmasterfree}) as
\begin{equation}
\begin{array}{l}
\Lambda_Z\langle X|B_Z^0\rangle=(1+\nu)^2 p^3\langle x-1,x,x+1|B^0_Z\rangle+q\langle x,x+1,x+2|B^0_Z\rangle+\\
\;\;\;\;\;\;\;\;\;\;\;\;\;\;\;\;\;\;
(1+\nu) p^2(q-\nu p)\langle x-1,x,x+2|B^0_Z\rangle+p(q-\nu p)\langle x-1,x+1,x+2|B^0_Z\rangle
\label{act}
\end{array}
\end{equation}
what corresponds exactly to the eigenproblem of operator $\mathbf{T}_{0}$ for the cluster $X=\{x,x+1,x+2\}$.

Using an induction, we can generalize (\ref{forbidden_3}) and (\ref{forbidden_4}):
\begin{equation}
\begin{array}{l}
\langle x,x+1,x+2,...,x+N-2,x+N-2|B^0_Z\rangle=\langle x,x+1,x+2,...,x+N-1|B^0_Z\rangle+\\
\frac{\nu^{N-1}p^{N-1}}{(1-p)^{N-1}}\big(\langle x-1,x,x+1,...,x+N-2|B^0_Z\rangle-
\langle x-1,x,x+1,...,x+N-3,x+N-1|B^0_Z\rangle\big), \label{8}
\end{array}
\end{equation}

\begin{equation}
\begin{array}{l}
\langle x,x,x+1,...,x+N-2|B^0_Z\rangle=\langle x,x+1,x+2,...,x+N-1|B^0_Z\rangle+\\
\frac{{\nu}p}{1-p}\Big(\langle x-1,x,x+1,...,x+N-2|B^0_Z\rangle-\langle x-1,x+1,x+2,...,x+N-1|B^0_Z\rangle\Big). \label{7}
\end{array}
\end{equation}

Suppose now that for $N$ particles we have
\begin{equation}
\begin{array}{l}
\sum\limits_{k_1=0}^{1}\sum\limits_{k_2=0}^{1}...\sum\limits_{k_{N}=0}^1p^{k_1}(1-p)^{1-k_1}p^{k_2}(1-p)^{1-k_2}...p^{k_{N}}(1-p)^{1-k_{N}}\times\\
\langle x-k_1,x+1-k_2,...,x+N-1-k_{N}|B^0_Z\rangle=(1+\nu)^{N-1}p^N\langle x-1,x,...,x+N-2|B^0_Z\rangle+\\
(1-p)\langle x,x+1,...,x+N-1|B^0_Z\rangle+\sum\limits_{k=1}^{N-1}(1+\nu)^{k-1}p^k(q-\nu{p})\times\\
\;\;\;\;\;\;\;\;\;\;\;\;\;\;\;\;\;\;\;\;\;
\;\;\;\;\;\;\;\;\;\;\;\;\;\;\;\;\;\;\;\;\;
\langle x-1,x,x+1,...,x+k-2,x+k,x+k+1,...,x+N-1|B^0_Z\rangle
\end{array}
\label{assumption}
\end{equation}
Then, for $N+1$ particles we obtain
\begin{equation}
\begin{array}{l}
\sum\limits_{k_1=0}^{1}\sum\limits_{k_2=0}^{1}...\sum\limits_{k_{N+1}=0}^1p^{k_1}(1-p)^{1-k_1}p^{k_2}(1-p)^{1-k_2}... p^{k_{N+1}}(1-p)^{1-k_{N+1}}\times\\
\;\;\;\;\;\;\;\;\;\;\;\;\;\;\;\;\;\;\;\;\;\;\;\;\;\;\;\;\;\;\;\;\;\;\;\;\;
\;\;\;\;\;\;\;\;\;\;\;\;\;\;\;\;\;\;\;\;\;\;\;\;\;\;\;\;\;\;\;\;\;\;\;\;\;
\langle x-k_1,x+1-k_2,...,x+N-k_{N+1}|B^0_Z\rangle=\\
\sum\limits_{k_{N+1}=0}^1p^{k_{N+1}}(1-p)^{1-k_{N+1}}\big((1+\nu)^{N-1}p^N\times
\langle x-1,x,...,x+N-2,x+N-k_{N+1}|B^0_Z\rangle+\\
\;\;\;\;\;\;\;\;\;\;(1-p)\langle x,x+1,...,x+N-1,x+N-k_{N+1}|B^0_Z\rangle+\sum\limits_{k=1}^{N}(1+\nu)^{k-1}p^k(q-\nu{p})\times\\
\;\;\;\;\;\;\;\;\;\;\;\;\;\;\;\;\;\;\;
\langle x-1,x,x+1,...,x+k,x+k+2,x+k+3,...,x+N-1,x+N-k_{N+1}|B^0_Z\rangle\big).
\end{array}
\end{equation}
Applying (\ref{8}),(\ref{7}) we derive
\begin{equation}
\begin{array}{l}
(1+\nu)^{N-1}p^N\big[p\langle x-1,x,...,x+N-2,x+N-1|B^0_Z\rangle+\\
\;\;\;\;\;\;\;\;\;\;\;\;\;\;\;\;\;\;\;\;\;\;\;\;\;\;\;\;\;\;\;\;\;\;\;\;
\;\;\;\;\;\;\;\;\;\;\;\;\;\;\;\;\;\;\;\;\;\;\;\;\;\;\;\;\;\;\;\;\;\;\;\;
(1-p)\langle x-1,x,...,x+N-2,x+N|B^0_Z\rangle\big]+\\
(1-p)\big[(1-p)\langle x,x+1,...,x+N-1,x+N|B^0_Z\rangle+p\langle x,x+1,x+2,...,x+N|B^0_Z\rangle+\\
\;\;\;\;\;\frac{\nu^{N}p^{N+1}}{(1-p)^{N}}(\langle x-1,x,x+1,...,x+N-1|B^0_Z\rangle-\langle x-1,x,x+1,...,x+N-2,x+N|B^0_Z\rangle)\big]+
\end{array}
\nonumber
\end{equation}
\begin{equation}
\begin{array}{l}
\sum\limits_{k=1}^{N}(1+\nu)^{k-1}p^k(q-\nu{p})\big[(1-p)\times\\
\;\;\;\;\;\;\;\;\;\;\;\;\;\;\;\;\;\;\;\;\;\;\;\;\;\;\;
\langle x-1,x,x+1,...,x+k-2,x+k,x+k+1,...,x+N-1,x+N|B^0_Z\rangle+\\
\;\;\;\;\;\;\;\;
\frac{1}{(1-p)^{N-k}}[\nu^{N-k}p^{N+1}(\langle x-1,x,x+1,...,x+N-1|B^0_Z\rangle+\\
\;\;\;\;\;\;\;\;\;\;\;\;\;\;\;\;
\frac{p}{1-p}\big((1-p)^{N+1-k}p^k\langle x-1,x,...,x+k-2,x+k,x+k+1...,x+N|B^0_Z\rangle)-\\
\;\;\;\;\;\;\;\;\;\;\;\;\;\;\;\;\;\;\;\;\;\;\;\;\;\;\;\;\;\;\;\;
\nu^{N-k}p^{N}(1-p)\langle x-1,x,x+1,...,x+N-2,x+N|B^0_Z\rangle\big)]\big]
\end{array}
\label{A9}
\end{equation}
The coefficient of the term $\langle x-1,x,...,x+N-2,x+N-1|B^0_Z\rangle$ is
\begin{equation}
\begin{array}{l}
(1+\nu)^{N-1}p^{N+1}+\frac{1}{(1-p)^{N-1}}(\nu^Np^{N+1})+\sum\limits_{k=1}^{N}(1+\nu)^{k-1}(q-\nu{p})\frac{1}{(1-p)^{N-k}}\nu^{N-k}p^{N+1}=\\
=p^{N+1}(1+\nu)^{N},
\end{array}
\label{10}
\end{equation}
The coefficient of the term $\langle x,x+1,...,x+N-1,x+N|B^0_Z\rangle$ is
\begin{eqnarray}
(1-p)^2+(1-p)\frac{1}{(1-p)^{N}}\frac{p}{1-p}(1-p)^{N+1}=1-p,
\label{11}
\end{eqnarray}
The coefficient of the term $\langle x-1,x,...,x+N-2,x+N|B^0_Z\rangle$ is
\begin{equation}
\begin{array}{l}
(1+\nu)^{N-1}p^{N}(1-p)-\frac{\nu^Np^{N+1}}{(1-p)^{N-1}}-p^{N+1}\sum\limits_{k=1}^{N-1}(1+\nu)^{k-1}(q-\nu{p})\frac{1}{(1-p)^{N-k}}\nu^{N-k}=\\
p^{N}(1+\nu)^{N-1}(q-{\nu}p)
\end{array}
\label{12}
\end{equation}
and the coefficient of the term $\langle x-1,x,...,x+k-2,x+k,x+k+1,...,x+N|B^0_Z\rangle$ is
\begin{eqnarray}
\frac{(1+\nu)^{k-1}}{(1-p)^{N-k}}(q-{\nu}p)p^k(1+\frac{p}{1-p})=\frac{(1+\nu)^{k-1}}{(1-p)^{N+1-k}}(q-{\nu}p)p^k \label{13}
\end{eqnarray}
Collecting all terms, we obtain for $N+1$ particles
\begin{equation}
\begin{array}{l}
\sum\limits_{k_1=0}^{1}\sum\limits_{k_2=0}^{1}...\sum\limits_{k_{N+1}=0}^1p^{k_1}(1-p)^{1-k_1}p^{k_2}(1-p)^{1-k_2}... p^{k_{N+1}}(1-p)^{1-k_{N+1}}\times\\
\;\;\;\;\;\;\;\;\;\;\;\;\;\;\;\;\;\;\;\;\;\;\;\;\;\;\;\;\;\;\;\;
\;\;\;\;\;\;\;\;\;\;\;\;\;\;\;\;\;\;\;\;\;\;\;\;\;\;\;\;\;\;\;\;
\langle x-k_1,x+1-k_2,...,x+N-k_{N}|B^0_Z\rangle= \\(1+\nu)^{N}p^{N-1}\langle x-1,x,...,x+N-1|B^0_Z\rangle+(1-p)\langle x,x+1,...,x+N|B^0_Z\rangle+\\
\;\;\;\;\sum\limits_{k=1}^{N}(1+\nu)^{k-1}p^k(q-\nu{p})\langle x-1,x,x+1,...,x+k-2,x+k,x+k+1,...,x+N|B^0_Z\rangle,
\end{array}
\end{equation}
q.e.d.

\end{document}